\documentclass[%
 reprint,
 superscriptaddress,
 amsmath,amssymb,
 aps,
]{revtex4-2}

\usepackage[mathlines]{lineno}

\usepackage{graphicx}
\usepackage{dcolumn}
\usepackage{multirow}
\usepackage{bm}
\usepackage{textcomp}
\usepackage{longtable}
\usepackage{booktabs}
\usepackage{mathrsfs}
\usepackage{epsfig}
\usepackage{rotating}
\usepackage{xfrac}
\usepackage{threeparttable}
\usepackage{lipsum} 
\newcommand{\mytilde}{{\raise.17ex\hbox{$\scriptstyle\mathtt{\sim}$}}}
\newcommand{\etal}{{\it et al.}}
\RequirePackage{hyperref}
\hypersetup
{
bookmarks=true,%
unicode=false,%
pdftoolbar=true,
pdfmenubar=true,
pdffitwindow=false,
pdfstartview=FitH,
pdftitle={PRL},
pdfauthor={LAM Yi Hua},
pdfsubject={Nuclear Astrophysics},
pdfcreator={LAM Yi Hua},
pdfproducer={LAM Yi Hua},
pdfkeywords={NucAstro},
bookmarksnumbered,%
colorlinks=true,%
linkcolor=blue,%
citecolor=magenta,%
filecolor=green,%
urlcolor=blue,
breaklinks, 
plainpages=false%
}
\usepackage{CJKutf8}

\newcommand{\cntextSimKai}[1]{\begin{CJK*}{UTF8}{gkai}#1\end{CJK*}}

\newcommand{\cntextTraKai}[1]{\begin{CJK*}{UTF8}{bkai}#1\end{CJK*}}
\newcommand{\jptextJap}[1]{\begin{CJK*}{UTF8}{min}#1\end{CJK*}}
\newcommand{\krtextKor}[1]{\begin{CJK*}{UTF8}{mj}#1\end{CJK*}}

\begin{document}

\preprint{APS/123-QED}

\title{Advancement of Photospheric Radius Expansion and Clocked Type-I X-Ray Burst Models with the New $^{22}$Mg$(\alpha,p)^{25}$Al Reaction Rate Determined at Gamow Energy}

\author{J. Hu (\cntextSimKai{胡钧})}
 \email{hujunbaggio@impcas.ac.cn}
 \affiliation{Institute of Modern Physics, Chinese Academy of Sciences, Lanzhou 730000, China.}
 \affiliation{School of Nuclear Science and Technology, University of Chinese Academy of Sciences, Beijing 100049, China}

\author{H. Yamaguchi (\jptextJap{山口\;英斉})}
 \affiliation{Center for Nuclear Study(CNS), the University of Tokyo, RIKEN campus, 2-1 Hirosawa, Wako, Saitama 351-0198, Japan}
 \affiliation{National Astronomical Observatory of Japan, 2-21-1 Osawa, Mitaka, Tokyo 181-8588, Japan}
%
\author{Y. H. Lam (\cntextTraKai{藍乙華})}
\email{lamyihua@impcas.ac.cn}
 \affiliation{Institute of Modern Physics, Chinese Academy of Sciences, Lanzhou 730000, China.}
 \affiliation{School of Nuclear Science and Technology, University of Chinese Academy of Sciences, Beijing 100049, China}
%
\author{A. Heger}
 \affiliation{School of Physics and Astronomy, Monash University, Victoria 3800, Australia}%
 \affiliation{OzGrav-Monash -- Monash Centre for Astrophysics, School of Physics and Astronomy, Monash University, VIC 3800, Australia}%
 \affiliation{Center of Excellence for Astrophysics in Three Dimensions (ASTRO-3D), Australia}
 \affiliation{The Joint Institute for Nuclear Astrophysics, Michigan State University, East Lansing, MI 48824, USA}%
%
\author{D.~Kahl}
 \affiliation{Extreme Light Infrastructure - Nuclear Physics, IFIN-HH, 077125 Bucharest-M\u{a}gurele, Romania}
 \affiliation{SUPA, School of Physics {\&} Astronomy, University of Edinburgh, Edinburgh EH9 3FD, United Kingdom}
%
\author{A.~M.~Jacobs}%
 \affiliation{The Joint Institute for Nuclear Astrophysics, Michigan State University, East Lansing, MI 48824, USA}%
 \affiliation{Department of Physics and Astronomy, Michigan State University, East Lansing, MI 48824, USA}
\author{Z. Johnston}%
 \affiliation{The Joint Institute for Nuclear Astrophysics, Michigan State University, East Lansing, MI 48824, USA}%
 \affiliation{Department of Physics and Astronomy, Michigan State University, East Lansing, MI 48824, USA}
%
\author{S.~W.~Xu~(\cntextSimKai{许世伟})}
\author{N.~T.~Zhang~(\cntextSimKai{张宁涛})}
\author{S.~B.~Ma~(\cntextSimKai{马少波})}
\author{L.~H.~Ru~(\cntextSimKai{茹龙辉})}
\author{E.~Q.~Liu~(\cntextSimKai{刘恩强})}
\author{T.~Liu~(\cntextSimKai{刘通})}
 \affiliation{Institute of Modern Physics, Chinese Academy of Sciences, Lanzhou 730000, China.}
\author{S. Hayakawa~(\jptextJap{早川\;勢也})}
\author{L. Yang~(\cntextSimKai{杨磊})}
\altaffiliation[Present address:]{ China Institute of Atomic Energy, P.O. Box 275(10), Beijing 102413, China}
\author{H.~Shimizu~(\jptextJap{清水\;英樹})}
 \affiliation{Center for Nuclear Study(CNS), the University of Tokyo, RIKEN campus, 2-1 Hirosawa, Wako, Saitama 351-0198, Japan}
\author{C.~B.~Hamill}
\author{A.~St\,J.~Murphy}
 \affiliation{SUPA, School of Physics {\&} Astronomy, University of Edinburgh, Edinburgh EH9 3FD, United Kingdom}
%
\author{J.~Su~(\cntextSimKai{苏俊})}
 \affiliation{College of Nuclear Science and Technology, Beijing Normal University, Beijing 100875, China}
\author{X. Fang~(\cntextSimKai{方晓})}
 \affiliation{Sino-French Institute of Nuclear Engineering and Technology, Sun Yat-Sen University, Zhuhai 519082, Guangdong, China}
\author{K.~Y.~Chae~(\krtextKor{채경육})}
\author{M.~S.~Kwag~(\krtextKor{곽민식})}
\author{S.~M.~Cha~(\krtextKor{차수미})}
\affiliation{Department of Physics, Sungkyunkwan University, Suwon 16419, Korea}
\author{N.~N.~Duy}
\affiliation{Department of Physics, Sungkyunkwan University, Suwon 16419, Korea}
\affiliation{Institute of Research and Development, Duy Tan University, Da Nang 550000, Vietnam}
\author{N.~K.~Uyen}
\author{D.~H.~Kim~(\krtextKor{김두현})}
 \affiliation{Department of Physics, Sungkyunkwan University, Suwon 16419, Korea}
\author{R. G. Pizzone}
 \affiliation{Laboratori Nazionali del Sud-INFN, Via S. Sofia 62, Catania 95123, Italy}
\author{M. La Cognata}
 \affiliation{Laboratori Nazionali del Sud-INFN, Via S. Sofia 62, Catania 95123, Italy}
\author{S. Cherubini}
 \affiliation{Laboratori Nazionali del Sud-INFN, Via S. Sofia 62, Catania 95123, Italy}
\author{S. Romano}
 \affiliation{Laboratori Nazionali del Sud-INFN, Via S. Sofia 62, Catania 95123, Italy}
 \affiliation{Dipartimento di Fisica e Astronomia ``Ettore Majorana'' - Università degli Studi di Catania}
 \affiliation{Centro Siciliano di Fisica Nucleare e Struttura della Materia (CSFNSM)}
\author{A.~Tumino}
 \affiliation{Laboratori Nazionali del Sud-INFN, Via S. Sofia 62, Catania 95123, Italy}
 \affiliation{Facolt{\`a} di Ingegneria e Architettura, Universit{\`a} degli Studi di Enna ``Kore'', Enna 94100, Italy}
\author{J.~Liang}
\author{A. Psaltis}
 \affiliation{Department of Physics {\&} Astronomy, McMaster University, Ontario L8S 4M1, Canada}
\author{M. Sferrazza}
 \affiliation{D\'{e}partement de Physique, Universit\'{e} Libre de Bruxelles, Bruxelles B-1050, Belgium}
\author{D. Kim~(\krtextKor{김다히})}
 \affiliation{Department of Physics, Ewha Womans University, Seoul 03760, Korea}
%
%
\author{Y. Y. Li (\cntextSimKai{李依阳})}
 \affiliation{Institute of Modern Physics, Chinese Academy of Sciences, Lanzhou 730000, China.}
 \affiliation{School of Nuclear Science and Technology, University of Chinese Academy of Sciences, Beijing 100049, China}
%
\author{S. Kubono (\jptextJap{久保野\;茂})}
 \affiliation{RIKEN Nishina Center, 2-1 Hirosawa, Wako, Saitama 351-0198, Japan}

\date{\today}

\begin{abstract}
We report the first (in)elastic scattering measurement of $^{25}\mathrm{Al}+p$ with the capability to select and measure in a broad energy range the proton resonances in $^{26}$Si contributing~to the $^{22}$Mg$(\alpha,p)$ reaction at type I x-ray burst energies.~We measured spin-parities of four resonances above the $\alpha$ threshold of~$^{26}$Si that are found to strongly impact the $^{22}$Mg$(\alpha,p)$ rate. The new rate advances a state-of-the-art model~to~remarkably~reproduce light curves~of the GS~1826$-$24 clocked burster~with~mean deviation~$<9$\% and permits us to discover a strong correlation between the He abundance in the accreting envelope of photospheric radius expansion burster and the dominance of $^{22}$Mg$(\alpha,p)$ branch.
\end{abstract}

\maketitle


Thermonuclear x-ray bursts (XRBs) are the most frequently recorded outbursts that happen in the Galaxy \cite{LEW93, SCH06, PAR13}. To date, 115 XRB sources have been discovered \cite{MINBAR}. More than 62 of the 115 sources categorized as photospheric radius expansion (PRE) bursters \cite{MINBAR} of which their bursting mechanism is still an unresolved puzzle due to their intricate hydrodynamics, e.g., the accretion-powered millisecond pulsar SAX J1808.4$-$3658 \cite{Wijnands1998,Chakrabarty2003}, which ignited the brightest XRB in recent history~\cite{Bult2019}. Its first multizone model was recently established~\cite{Johnston2018,Goodwin2019} and is subject to verification; conversely, it offers a first concurrent sensitivity study on reaction rates for the light curves, fluences, and recurrence times, especially the competition between important reactions at a branching point during the onset of an XRB. 
The GS~1826$-$24 clocked burster~\cite{Makino1988,Tanaka1989,Ubertini1999} is the most investigated due to its nearly consistent accretion rate and light-curve shape. Its XRB serves as a laboratory to probe 
the \emph{rp}-process path~\cite{Schatz1998,Fisker2008},
compactness \cite{Randhawa2020}, and equation of state of the accreting neutron star \cite{STE10,Dohi2020}.
Thus, the best model describing the GS~1826$-$24 light curves is highly desired within the community. The first quantitative comparison of its modeled and observed light curves could only be achieved 19 yr after its discovery~\cite{Heger2007}; however, up to now, the modeled burst tail does not exactly conform with observation; a similar problem also occurs in other multizone models~\cite{MESA2015,Meisel2018a,Randhawa2020}. 
It is crucial to verify whether the incapability of the model is due to astrophysical configurations or some influential nuclear reaction rates.




Two recent sensitivity studies performed by \citet{Cyburt2016} and by \citet{Jacobs2020} using GS~1826$-$24 models \cite{Heger2007} reveal that the $^{22}$Mg($\alpha$, $p$) rate is the most decisive $\alpha p$-process reaction in $sd$-shell nuclei influencing burst light curves, see Supplemental Material (SM) \citep{SupplementMaterial}.
The $^{22}$Mg$(\alpha,p)$ rate proposed by the compilation reaction library REACLIB v2.2~\citep{Cyburt2010}, however, is generated using the Hauser-Feshbach (HF) model \cite{NONSMOKER} assuming a rather high level-density of $^{26}$Si. This assumption may be invalid and inapplicable considering the selectivity of the $(\alpha,p)$ reaction for natural parity states; moreover, the rate from a high resolution $^{28}$Si$(p,t)^{26}$Si measurement~\cite{MAT11} was deduced without the experimental information of important resonances within the Gamow window, resulting in a rate up to 6 orders of magnitude lower than the HF-model $^{22}$Mg$(\alpha,p)$ rate. %
Recently, the first direct measurement of the $^{22}$Mg$(\alpha, p)$ reaction was performed by \citet{Randhawa2020}. The evaluated $^{22}$Mg$(\alpha, p)$ rate is, however, based on a rather low $^{22}$Mg beam intensity of $\mytilde$900~pps which did not permit a direct measurement of $^{22}$Mg$(\alpha, p)$ reaction in the Gamow window of XRBs. Only protons with a limited range (90$^\circ$--120$^\circ$) were analyzed and the \texttt{PACE4} code~\citep{LISE} had to be used to simulate the total cross section. Consequently, they only obtained cross sections corresponding to 2.6~GK.
The reaction rates at XRB temperatures (0.7--1.0~GK) were then extrapolated relying on the \texttt{TALYS} code, without direct experimental information at the relevant temperature. 
Such an extrapolation could induce a large additional uncertainty that was not presented in Ref.~\citep{Randhawa2020}.
Thus, confirming the $^{22}$Mg$(\alpha,p)$ rate with precisely measured resonance properties within Gamow window of low uncertainty is crucial to regulate better XRB models to unfold the physics of accreting neutron stars.

In this Letter, we report the first measurement of $^{25}\mathrm{Al}+p$ (in)elastic scattering at x-ray burst energies to deduce the $^{22}$Mg$(\alpha, p)^{25}$Al rate. This technique overcomes the difficulties in direct measurement due to the low-cross-section nature of $^{22}$Mg$(\alpha, p)$ reaction in the Gamow window. We used the radioactive ion beam separator (CRIB) \cite{YAN05,Kubono2002,Yamaguchi2020} 
of the University of Tokyo. 
A primary beam of $^{24}$Mg$^{8+}$ at 8.0 MeV/nucleon 
and 1~e$\mu$A bombarded a cryogenic D$_2$ target \cite{YAM08} to produce a secondary beam of $^{25}$Al.
The $^{25}$Al beam was purified by CRIB using the in-flight method. The $^{25}$Al beam, with an energy of $142\pm1$~MeV and an average intensity of $2.0\times10^5$~pps, was then delivered to the F3 experimental scattering chamber and bombarded a 150-$\mu$m-thick CH$_2$ target,
similarly to Ref.~\cite{YAM09}.

The beam particles were identified 
event by event and the $^{25}$Al beam purity was typically 70\%.
The impurity was mostly $^{24}$Mg, clearly discriminated by the timing information.



The recoiling protons were measured using three sets of silicon detector telescopes at central angles of $\theta_{\mathrm{lab}}=0^{\circ}$, $20^{\circ}$, and $23^{\circ}$. Each telescope consisted of a 65-$\mu$m-thick and double-sided ($16\times16$ strips) silicon detector and two 1500-$\mu$m-thick pad detectors. Protons were clearly identified from other light ions with the $\Delta E$$-$$E$ method. To identify the inelastic contribution, an array of ten NaI detectors was mounted immediately above the target to detect the $\gamma$ rays from the decay of excited states of $^{25}$Al. Each NaI detector with a geometry of $50\times50\times100$~mm, with the arrary covering 20\% of the total solid angle. These detectors had an average energy resolution of 13.5\% in full width at half maximum~(FWHM)~for~662-keV~$\gamma$~rays. 
In addition, an 80-$\mu$m-thick carbon target was used in a separate run for subtracting the carbon background contribution. 

The $E_{c.m.}$ resolution of the excitation function was 30-90 keV (FWHM), depending on the energy, for the Si telescope around $\theta_{\mathrm{lab}}=0^{\circ}$. The uncertainty was mostly from energy straggling of the particles in the thick target, along with the energy resolution of the silicon detectors. At larger angles, the angular resolution of the recoiling proton produced a larger energy uncertainty and the resulting energy resolution was 75$-$200~keV at $\theta_{\mathrm{lab}}$ $\mytilde20^{\circ}$. In this Letter, we focus on the forward angle measurement, where we had the highest resolution to determine the resonance parameters.

\begin{figure}[t!]
\centering
\includegraphics[width=7.7cm]{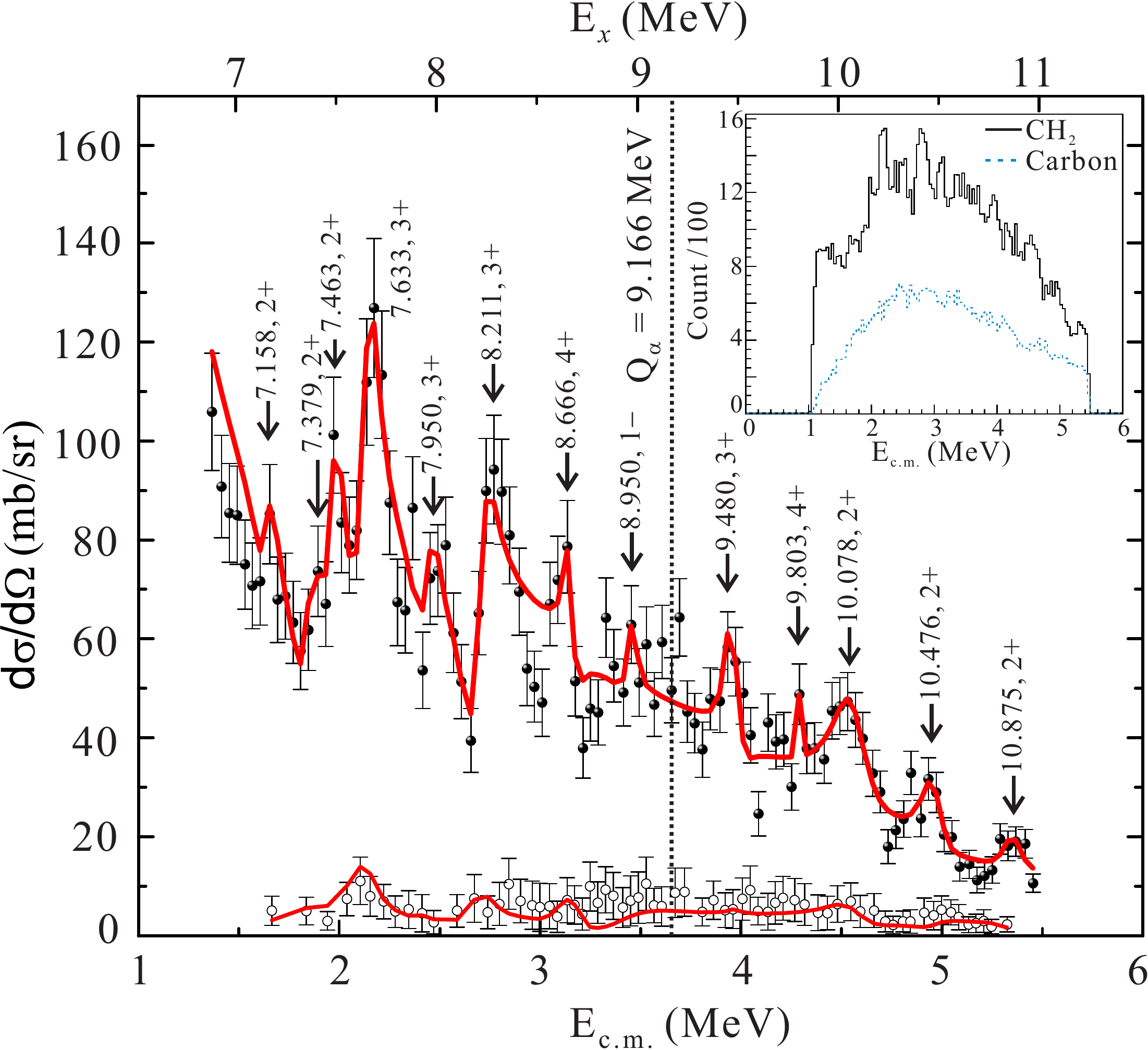}
\vspace{-3mm}
\caption{\label{fig3}Excitation function of $^{25}\mathrm{Al}+p$ elastic scattering at $\theta_{\mathrm{lab}}=0$~--~$8^{\circ}$. Elastic scattering data (filled circles); inelastic scattering data (open circles);
the best R-matrix fit (red curve); the $\alpha$ threshold (dotted line); Inset: the CH$_2$ spectrum with the normalized carbon background.}
\vspace{-8mm}
\end{figure}

The excitation function of $^{25}\mathrm{Al}+p$ elastic scattering has been deduced using the standard procedure as described in Refs.~\cite{YAM09, TER03, TER07, HE07}. The cross section of inelastic scattering, less than 12\% of the elastic scattering, was deduced by analyzing gamma-coincident events as plotted in Fig.~\ref{fig3}, and its contribution was subtracted from the total excitation function. The excitation function~around $\theta_{\mathrm{lab}}=0^{\circ}$ is shown in Fig.~\ref{fig3}. Several resonances are clearly evident in the spectrum. To determine the parameters of observed resonances, R-matrix calculations have been performed using \texttt{AZURE2}~\cite{AZU10} with a channel radius of $R=1.4\times(1+25^{1/3})$~fm for the $^{25}\mathrm{Al}+p$ system.

The ground-state spin-parity configurations of $^{25}$Al and proton are 5/2$^+$ and 1/2$^+$, respectively. Thirteen resonances have been analyzed, and the best fit curve is shown in Fig.~\ref{fig3}. The resonance properties are listed in Table~\ref{tab1}. The lowest five states are in good agreement with the previous $^{25}\mathrm{Al}+p$ resonant scattering measurements \cite{CHE12, JUN14}, except the weak 7.379-MeV resonance, where our $\Gamma_{p0}$ is larger than theirs and the 4$^+$ assignment by Jung {\it et al.} \cite{JUN14} cannot reproduce the present data well. The resonances at 8.211 and 8.666 MeV may correspond to the ones observed in Ref.~\cite{MAT10}, and a spin-parity of 1$^-$ was assigned to the 8.211-MeV resonance based on the mirror assignment. Our analysis shows the assignment as $1^-$ strongly disagrees with our data, however, whereas $3^+$ best matches our data.
\citet{BOH82} also discovered the 8.666-MeV state via a $^{24}$Mg$(^3\mathrm{He},n)^{26}$Si measurement and a tentative $J^\pi$ assignment (1$^-$ or 2$^+$) was made based on a distorted wave Born approximation (DWBA) calculation. Our fitting result disagrees with theirs, but supports the 4$^+$ assignment made by \citet{MAT10}. Although higher resonances had been observed by previous studies \cite{SHI05, THO04, MAT11}, no $J^\pi$ was determined. We observed these resonances in the present work, and assigned their tentative $J^\pi$ with our best R-matrix fit ($\chi^2/$DOF $=1.08$ for 103 DOF). 
Our presently assigned spin parities generally agree with known states of $^{26}$Si.
Taking into account all possible assignments for the 9.480-, 9.803-, and 10.078-MeV states, the total $^{22}$Mg$(\alpha,p)$ rate changes up to a factor of 0.44 for temperature above 0.7~GK. The minimum $\chi^2$ of the R-Matrix fit supports the 10.476-MeV state to be assigned as 2$^+$. This state can also be produced via ($p$,$t$) reaction \cite{MAT11} which preferentially excites natural-parity states. The 10.875-MeV state can only be either 2$^+$, 3$^+$, or 4$^+$ due to the selection rule of Gamow-Teller transitions~\cite{THO04}. 
We assign a 2$^+$ to the 10.875-MeV state, which gives the minimum $\chi^2$. However, the assignments of 3$^+$ and 4$^+$ only produce deviations in $\chi^2$ within the standard deviation $\sigma$ (0.50$\sigma$ and 0.62$\sigma$, respectively), and thus we also consider its possibility as 3$^+$ or 4$^+$ in the analysis below as it determines the rate above 1~GK. Further information of the R-matrix analysis is detailed in the SM \citep{SupplementMaterial}.
To constrain the level properties of the states contributing the reaction rate, we also performed a simultaneous fit for both elastic and inelastic scattering data. With the limited data quality, we obtained the upper limits of inelastic proton widths, $\Gamma_{p1,\mathrm{max}}$ (Table~\ref{Tab2}).

\selectfont
\begin{table*}[htp]
\footnotesize
\begin{threeparttable}[b]
\caption{\label{tab1}
The presently determined energy levels of $^{26}$Si compared with literature.}
\begin{tabular*}{\textwidth}{c@{\extracolsep{\fill}}lcccccc}
\toprule
\midrule
\multicolumn{1}{c}{} &\multicolumn{3}{c}{$^{26}$Si present work} & \multicolumn{3}{c}{$^{26}$Si from other works} & \\
\cmidrule(lr){2-4}  \cmidrule(lr){5-8} 
No. & E$_x$ (MeV)\tnote{a} & $J^{\pi}$ & $\Gamma_{p0}$\tnote{b} (keV) & E$_x$ (MeV) & $J^{\pi}$ & $\Gamma_{p0}$ (keV) & Refs.\\
\midrule
  1. & 7.158(13)  & 2$^+$ & 6(3)    & 7.162(14) / 7.147(27) & 2$^+$                    & 7(4) / 2.7(1)             & \cite{CHE12} / \cite{JUN14}\\
  2. & 7.379(18)  & 2$^+$ & 28(14)   & 7.402(40) / 7.401(28) & 2$^+$ / 4$^+$            & 6(4) / 1.1(1)            & \cite{CHE12} / \cite{JUN14}\\
  3. & 7.463(18)  & 2$^+$ & 51(9)   & 7.484(13) / 7.484(28) & 2$^+$                    & 46(11) / 15.9(3)          & \cite{CHE12} / \cite{JUN14}\\
  4. & 7.633(20)  & 3$^+$ & 46(8)   & 7.704(13) / 7.654(29) & 3$^+$ / (2$^+$, 3$^+$)   & 41(6) / (30.1(5), 19.5(3))& \cite{CHE12} / \cite{JUN14}\\
  5. & 7.950(22)  & 3$^+$ & 10(5)   & 8.015(14) / 7.977(30) & 3$^+$ / (2$^+$, 3$^+$)   & 15(5) /  (4.5(3), 3.6(2)) & \cite{CHE12} / \cite{JUN14}\\
  6. & 8.211(24)  & 3$^+$ & 48(10)  & 8.222(5)              & 1$^-$                    &                           & \cite{MAT10} \\
  7. & 8.666(25)  & 4$^+$ & 8(5)    & 8.700(30) / 8.687(12) & (1$^-$, 2$^+$) / (4$^+$) &                           & \cite{MAT10} / \cite{BOH82} \\
  8. & 8.950(30)  & 1$^-$ & 16(5)    & 8.952(7)              &                          &                           & \cite{SHI05}\\
  9. & 9.480(30)  & 3$^+$ & 15(4)   & 9.433(4)              &                          &                           & \cite{THO04} \\
 10. & 9.803(32)  & 4$^+$ & 2(1)    & 9.802(7)              &                          &                           & \cite{SHI05} \\
 11. & 10.078(36)\tnote{c} & 2$^+$ & 164(30) & 10.070(8)             &                          &                           & \cite{SHI05} \\
 12. & 10.476(40) & 2$^+$ & 54(22)  & 10.436(10)            &                          &                           & \cite{MAT11} \\
 13. & 10.875(45) & 2$^+$ & 57(21)  & 10.827(8)             &                          &                           & \cite{THO04} \\
 \bottomrule
\end{tabular*}
\begin{tablenotes}
 \footnotesize
 \item[a] Statistical errors due to the R-matrix fit folded with systematic uncertainty of 12--35~keV is given in parentheses.
 \item[b] Elastic scattering proton widths.
 \item[c] An 1$^+$ assignment is not excluded, but not preferred from the inelastic data and its influence on the final reaction rate is negligible.
\end{tablenotes}
\end{threeparttable}
\vspace{-8mm}
\end{table*}

\begin{figure}[t]
\vspace{-5mm}
\centering
\includegraphics[width=8.0cm]{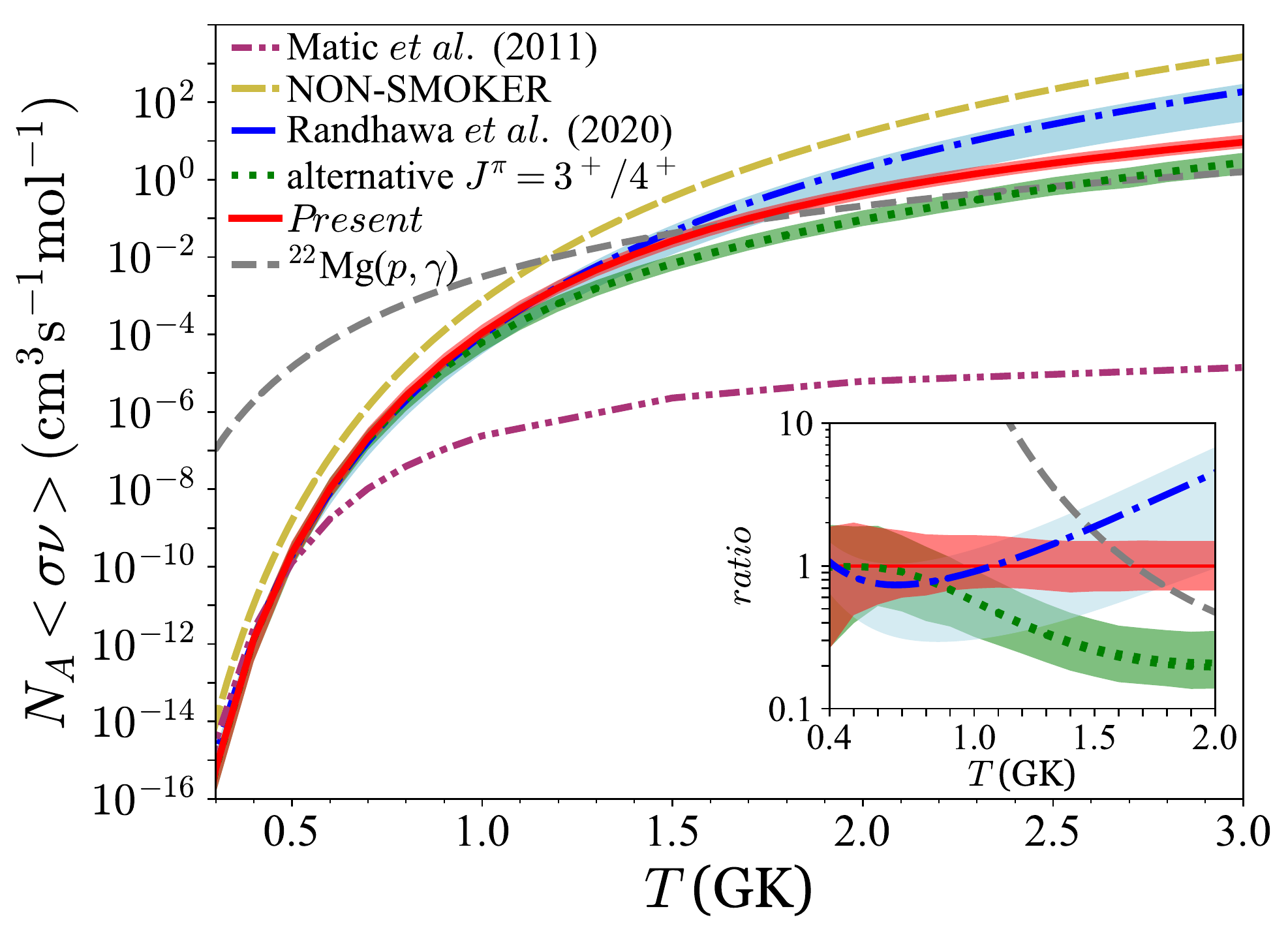}
\vspace{-4mm}
\caption{\label{fig4} The $^{22}$Mg$(\alpha,p)$ rates. The uncertainty of the present rate (red zone) is estimated via Monte-Carlo calculation \citep{Thompson1999} considering all errors from the present experimental measurement.
Both possible rates with $J^\pi$(10.875 MeV)=3$^+$ or 4$^+$ are not distinguishable, plotted as a green line and labeled as ``alternative $J^\pi$=3$^+$/4$^+$''. \citet{Randhawa2020} rate uncertainty~is~the~blue zone. Inset: the ratios of Randhawa \etal, or ``alternative $J^\pi$=3$^+$/4$^+$'' or $^{22}$Mg$(p,\gamma)$~\citep{Cyburt2010,Iliadis2010} rate to the present $^{22}$Mg$(\alpha,p)$ rate.}
\vspace{-7mm}
\end{figure}

The $^{26}$Si levels above the $\alpha$ threshold are expected to characterize the $^{22}$Mg$(\alpha,p)$ rates. 
As the widths are broad for the 10.078-, 10.476-, and 10.875-MeV states, we applied the broad-resonance approximation, in which the reaction rates can be obtained from~\cite{Iliadis2007},\vspace{-3.5mm}

\begin{eqnarray}
N_A\langle\sigma\upsilon\rangle =&& \sqrt{2\pi}\frac{N_A\hslash^2}{(\mu kT)^{3/2}}\sum_i\omega_i\int_0^\infty e^{-E/kT}\\
&&\times\frac{\Gamma_\alpha(E)\Gamma_p(E+Q)}{(E-E_R^i)^2+\Gamma(E)^2/4}dE \,[\mathrm{cm}^{3}\mathrm{s}^{-1}\mathrm{mol}^{-1}]\, . \nonumber
\end{eqnarray}
\vspace{-0.5mm}%
Here, $\mu$ is the reduced mass of the target and projectile, $T$ is the temperature, E$_R$ is the energy of the resonance, and the statistical factor $\omega$=$2J_i+1$. The energy dependence of the widths was taken into account by letting the partial widths $\Gamma_\alpha$ and $\Gamma_p$ vary as,  
$
 \Gamma_x^i(E)=\Gamma_x^i(E_R^i)\left[P_\ell(E)/P_\ell(E_R^i)\right]
$
where the $P_\ell$ are the Coulomb penetrabilities for the $\alpha$ and $p$ channels, respectively. The partial width $\Gamma_p(E_R)$ is from our R-matrix fit, and $\Gamma_\alpha(E_R)$ can be inferred from the mirror nucleus $^{26}$Mg via the isospin symmetry relation, $\Gamma_\alpha^i=C^2S_\alpha\Gamma_\alpha^{i,\mathrm{SP}}$, where the $C^2S_\alpha$ is the $\alpha$-spectroscopic factor and $\Gamma_\alpha^{\mathrm{SP}}$ is the single-particle $\alpha$ width. 
We adopted the average $C^2S_\alpha$ values from Ref.~\cite{MAT11}; $C^2S_\alpha(4^+)=0.015$ and 
$C^2S_\alpha(2^+)=0.037$, with uncertainties of a factor of 2, as in \cite{Nesaraja2007}. Table~\ref{Tab2} shows the adopted resonance parameters in obtaining the $^{22}$Mg$(\alpha,p)$ rates, which are shown together with the rates from the HF model (hereinafter \texttt{NON-SMOKER})~\cite{NONSMOKER} and~\citet{MAT11} in Fig.~\ref{fig4}. 
The resonance $J^\pi(10.875$~MeV$)=3^+$ does not contribute to the $^{22}$Mg$(\alpha,p)$ rate whereas the contribution from assuming it as 4$^+$ is much lower than assuming it as 2$^+$. Both possible $^{22}$Mg$(\alpha,p)$ rates assuming $J^\pi(10.875$~MeV$)=3^+$ or $4^+$ are similar and the difference in reaction rate is only up to a factor of 0.27.  Note that in the critical temperature range for XRB ignition, the \texttt{NON-SMOKER} $^{22}$Mg$(\alpha,p)$ rate differs from ours by a factor of $\mytilde$10 from $\mytilde0.4$ to $\mytilde1$~GK, and varies up to a factor of $\mytilde$160 at 3~GK. Because of the missing resonance data of $^{26}$Si above 10~MeV excitation energy in~\citet{MAT11}, there is a discrepancy of about 1 to 5 orders of magnitude between our new rate and the Matic \etal~rate for $T=0.7$~--~$3$~GK (Fig.~\ref{fig4}). The $^{22}$Mg$(\alpha,p)$ rate by \citet{Randhawa2020} approximated with the \texttt{NON-SMOKER} $^{22}$Mg$(\alpha,p)$ rate divided by 8, is also shown in Fig.~\ref{fig4}. 
Although their evaluated rate does not largely deviate from our present rate at around 1~GK and below, we caution that their evaluation may underestimate the uncertainty due to the theoretical extrapolation without considering each resonance explicitly. Our $^{22}$Mg$(\alpha,p)$ rate has a significantly lower uncertainty than theirs (Fig.~\ref{fig4}) even if such possible underestimation is ignored, see SM \citep{SupplementMaterial} for the further error estimation.
Our final rate is merely enhanced by at most 10\% when considering the additional $\Gamma_{p1,\mathrm{max}}$.

\begin{table}[tp]
\footnotesize
  \begin{threeparttable}[t]
  \caption{\label{Tab2}Resonance parameters for the $^{22}$Mg$(\alpha,p)$ rates.}
  \begin{tabular*}{\columnwidth}{c@{\extracolsep{\fill}}cccc}
   \toprule
   \midrule
   $E_x$ (MeV) & $J^{\pi}$ & $\Gamma_{\alpha}$ (eV) & $\Gamma_{p0}$ (keV) & $\Gamma_{p1,\mathrm{max}}$ (keV) \\
   \hline
   9.803(32) & 4$^+$ & 9.69 $\times10^{-13}$ & 2(1)    & 5.9 $\times10^{-3}$\\
  10.078(36) & 2$^+$ & 1.13 $\times10^{-6}$  & 164(30) & 22.6 \\
  10.476(40) & 2$^+$ & 1.80 $\times10^{-3}$  & 54(22)  & 9.9  \\
  10.875(45) & 2$^+$ & 1.70 $\times10^{-1}$  & 57(21)  & 1.0  \\
  \bottomrule
  \end{tabular*}
\end{threeparttable}
\end{table}

\begin{figure}[t!]
\begin{center}
\includegraphics[width=9.3 cm, angle=0]{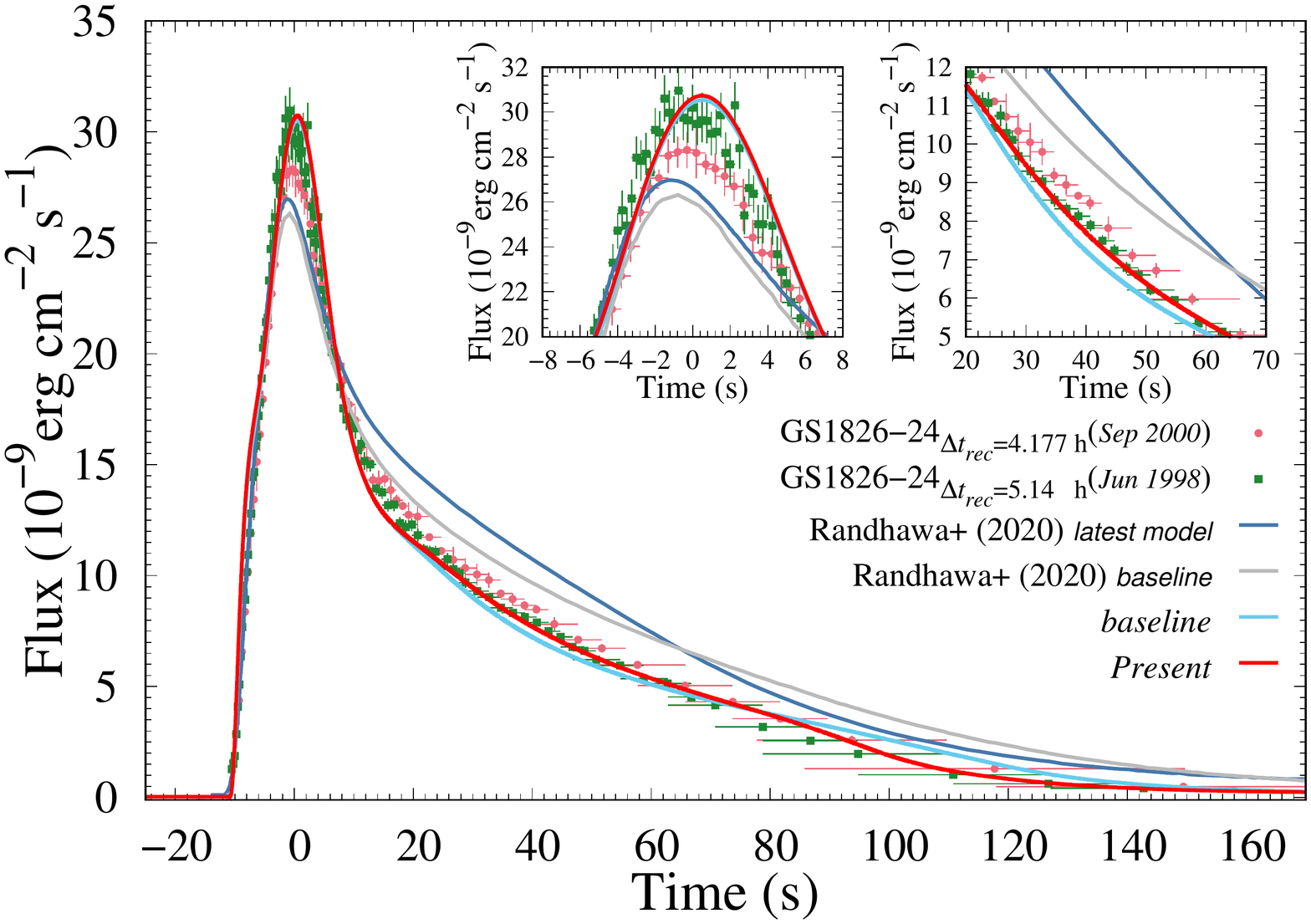}
\vspace{-7mm}
\caption{\label{fig:GS1826_Flux}The best fit \emph{baseline} and \emph{Present} modeled light curves to the observed light curve of epoch \emph{Jun 1998}, and the best fit \citet{Randhawa2020} light curves to epoch \emph{Sep 2000}. The magnified light curves at the burst peak and $t=20$~--~$70$~s are shown in the left and right insets, respectively.
}
\end{center}
\vspace{-9mm}
\end{figure}

\emph{GS 1826$-$24 clocked burster:} To quantitatively compare with the GS 1826$-$24 burster (Fig.~\ref{fig:GS1826_Flux}), we adopt the best fit model from~\citet{Jacobs2020}, which has~a ratio of accreted $^1$H to $^4$He of 2.39, a Carbon-Nitrogen-Oxygen (CNO) metal mass fraction of 0.0075, and an accretion rate of 3.325$\times10^{-9}~M_{\odot}$yr$^{-1}$, as our \emph{baseline} model. We update it with the present $^{22}$Mg$(\alpha,p)$ rate to represent the \emph{Present} model. The generated burst luminosity, $L_x$, by the 1D multizone hydrodynamic \texttt{KEPLER} code~\citep{Woosley2004,Heger2007} is related to observational flux, $F_x$ by scaling with $\left[4\pi d^2\xi_b(1+z)^2\right]^{-1}$~\citep{Johnston2020}, where $d$ is the distance, $\xi_b$ incorporates the possible burst-emission anisotropy, and the redshift, $z$, expands the light curve when transforming into an observer's frame. Instead of specifically selecting data  close to the burst peak at $t=-10$~to~$40$~s~\citep{Randhawa2020,Meisel2018a}, we impartially select all observational data of the entire burst timespan to fit our modeled bursts. The modeled bursts are averaged and fitted to the averaged light curve of GS~1826$-$24 epoch \emph{Jun 1998}~\citep{Galloway2017}, which were recorded by the Rossi X-ray Timing Explorer (RXTE) Proportional Counter Array~\citep{Galloway2004,Galloway2008,MINBAR}. 



The \emph{baseline} light curve at $t=16$~--~$76$~s is enhanced and the discrepancy with observed data becomes only up to 6\% due to the present and lower $^{22}$Mg$(\alpha,p)$ rate, which at low temperature competes with $^{22}$Mg$(\beta\nu)$ decay and overcomes $^{22}$Mg$(p,\gamma)$ at higher temperature $T>1.67^{+0.15}_{-0.13}$~GK instead of at $T>1.16$~GK compared to the \texttt{NON-SMOKER} $^{22}$Mg$(\alpha,p)$ rate (Fig.~\ref{fig4}). The alternative $J^\pi=3^+/4^+$ rate yields only 3\% deviation from the observed data at $t$=16--76~s, which is not discernible in Fig.~\ref{fig:GS1826_Flux}. The matter flow is more~siphoned~out~to $^{22}$Mg$(p,\gamma)^{23}\mathrm{Al}(p,\gamma)^{24}\mathrm{Si}(\alpha,p)$, enriching more proton-rich nuclei nearer to dripline past the \emph{sd}-shell. These nuclei burn hydrogen after the burst peak and enhance the light curve at $t=16$~--~$76$~s, depleting  hydrogen that is to be burnt by further $(p,\gamma)$ reactions at later time $t=80$~--~$150$~s. Hence, the observed light curve profile at $t=80$~--~$150$~s is noticeably reproduced. Therefore, the present work experimentally validates the predicted light curve trend in Ref.~\cite{Cyburt2016} and enhances a state-of-the-art model to remarkably reproduce the GS~1826$-$24 light curve with mean deviation $<9$~\%, see SM \citep{SupplementMaterial}. In the latest model by~\citet{Randhawa2020} (the blue line in Fig.~\ref{fig:GS1826_Flux}), a similar trend is manifested at $t=8$~--~$64$~s, however, it deviates their baseline model farther away from observation and affects their fitted redshift-distance. 

\begin{figure}[t!]
\begin{center}
\includegraphics[width=8.7cm, angle=0]{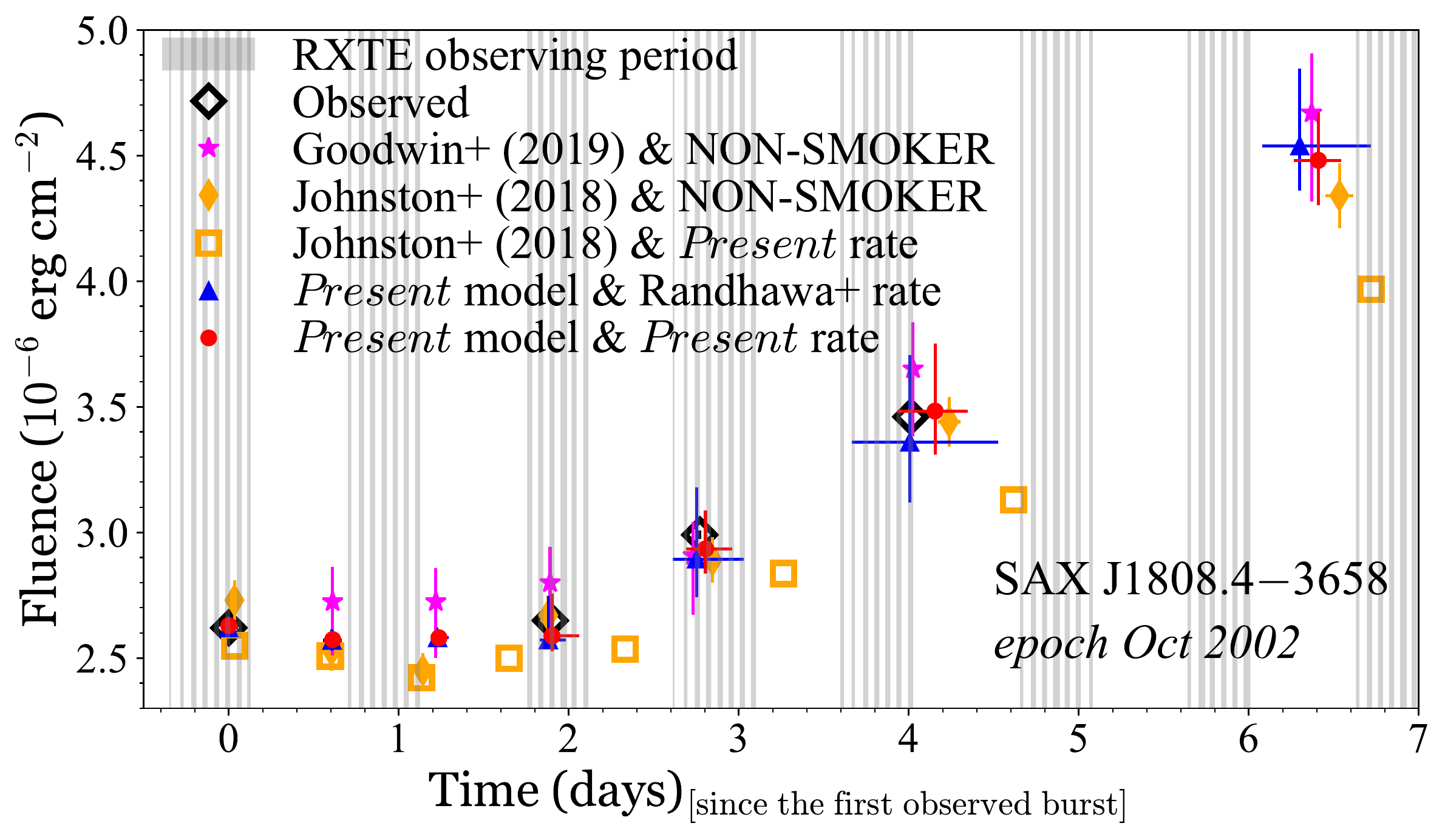}
\vspace{-3mm}
\caption{\label{fig:fluence}The bursts' fluences (integration of flux over time) and times for SAX J1808.4$-$3658 burster, based on the RXTE observation~\cite{MINBAR}, Johnston \etal~\cite{Johnston2018} and Goodwin \etal~\cite{Goodwin2019} models, and present calculations. Johnston \etal~\cite{Johnston2018} model is adopted to study the present
and Randhawa \etal~rates.}
\end{center}
\vspace{-7mm}
\end{figure}

\emph{SAX J1808.4$-$3658 PRE burster:} The initial good-fit SAX J1808.4$-$3658 PRE models constructed by \citet{Johnston2018} and studied by \citet{Goodwin2019} are based on the \texttt{KEPLER} code using the \texttt{NON-SMOKER} $^{22}$Mg$(\alpha,p)$ rate but these models can still provide us a unique and sensitive study for competition between the $^{22}$Mg$(\alpha,p)$ and $^{22}$Mg$(p,\gamma)$ reactions because the temperature of competition between both reactions, $T_C$ (the intersection of $^{22}$Mg$(\alpha,p)$ and $^{22}$Mg$(p,\gamma)$ \citep{Cyburt2010,Iliadis2010} rates in inset of Fig.~\ref{fig4}), is within the range of accreting-envelope maximum temperature, 1.1$\leqslant$$T_\mathrm{max}$/GK$\leqslant$1.6, during a typical PRE burst, and the He and H abundances are almost equal in the accreting envelope of SAX J1808.4$-$3658 PRE burster \citep{Johnston2018,Goodwin2019}. 
The present $^{22}$Mg$(\alpha,p)$ rate which has the lowest uncertainty among all available rates precisely locates the $T_C=1.67^{+0.15}_{-0.13}$~GK constricting the $^{22}$Mg$(\alpha,p)$ branch. With our new rate, the previous model parameters do no longer well reproduce the observation (orange squares in Fig.~\ref{fig:fluence}). With only constraining the He abundance in the accreting envelope to be $X_\mathrm{He}=56.7\pm0.3$\%, we successfully regulated the $^{22}$Mg$(\alpha,p)$ and $^{22}$Mg$(p,\gamma)$ branches and improved the modeled fluences closer to observation (red dots in Fig.~\ref{fig:fluence}). The He-abundance constraint reveals a strong correlation with the dominance of $^{22}$Mg$(\alpha,p)$ branch and introduces a striking advancement for the pioneering PRE model. The approximated $^{22}$Mg$(\alpha,p)$ rate~\cite{Randhawa2020} with large uncertainty, however, estimates a wide range of $T_C=1.4$~--~$1.8$~GK; also the propagation of their rate uncertainty yields a less constrained range of He abundance $X_\mathrm{He}=56.1\pm1.1$\% causing large uncertainty in fluences and times (blue triangles in Fig.~\ref{fig:fluence}). 

In summary, we have performed the first (in)elastic scattering measurement of $^{25}\mathrm{Al}+p$~with~the~capability to select and measure proton resonances contributing to the $^{22}$Mg$(\alpha,p)^{25}$Al reaction at XRB temperature. This provides the spectroscopic information of four resonances above the ${\alpha}$ threshold of $^{26}$Si that strongly influence the $^{22}$Mg$(\alpha,p)^{25}$Al reaction rate.~We successfully deduced the $^{22}$Mg$(\alpha,p)^{25}$Al rate via experiment without~implementing a scaling factor on a Hauser-Feshbach statistical model rate as was done in Ref.~\citep{Randhawa2020}. The improved nuclear physics input permits us to better reproduce the observed GS 1826$-$24 light curves than the previous model (see SM \citep{SupplementMaterial}) and to further constrain the SAX J1808.4$-$3658 model.


This experiment was performed at the RI Beam Factory operated by RIKEN Nishina Center and CNS, University of Tokyo.~We would like to thank the CRIB and RIKEN accelerator staffs for their dedication to~this project. We thank D. K. Galloway~for stimulating discussions and help in the comparison~with GS~1826$-$24 and SAX J1808.4$-$3658 XRB sources.
This work is financially supported by the Major State Basic Research Development Program of China (2016YFA0400503, 2016YFA0400501, and 2016YFA0400504), the Strategic Priority Research Program of Chinese Academy of Sciences, Grant No. XDB34020204, JSPS KAKENHI (Nos. 16K05369, No. 19K03883, and No. 18K13556), the Ministry of Education, Culture, Sports, Science and Technology (MEXT) of Japan, Chinese Academy of Sciences President's International Fellowship Initiative (No. 2019FYM0002). Y.H.L., K.Y.C., X.F. thank the National Natural Science Foundation of China (No. 11775277, No. 31211775277, and No. 11805291). 
The Edinburgh group is appreciative of funding from the UK STFC. 
A.H. is supported by the Australian Research Council Centre of Excellence for Gravitational Wave Discovery (OzGrav, No. CE170100004) and for All Sky Astrophysics in 3 Dimensions (ASTRO 3D, No. CE170100013). 
A.H., A.M.J, and Z.J. acknowledge support from the US National Science Foundation under Grant No. PHY-1430152 (JINA - Center for the Evolution of the Elements).
K.Y.C. was supported by National Research Foundation of Korea (Nos. 2020R1A2C1005981, 2019K2A9A2A10018827, and 2016R1A5A1013277). M.S. is supported from Fonds de la Recherche Scientifique - FNRS Grant Number 4.45.10.08. 
Y.H.L. and A.H. deeply appreciate the computing resources provided by the Institute of Physics (PHYS\_T3 cluster) and the ASGC (Academia Sinica Grid-computing Center) Distributed Cloud resources (QDR4 cluster) of Academia Sinica, Taiwan, and Gansu Advanced Computing Center.

\nocite{*}
\providecommand{\noopsort}[1]{}\providecommand{\singleletter}[1]{#1}%
%

\newpage
\begin{center}
{\bf \large Supplemental Material}
\label{sec:SM}
\end{center}
\begin{center}
{\bf More details of the R-matrix analysis}
\end{center}

\begin{figure}[t!]
\begin{center}
\includegraphics[width=8.6 cm]{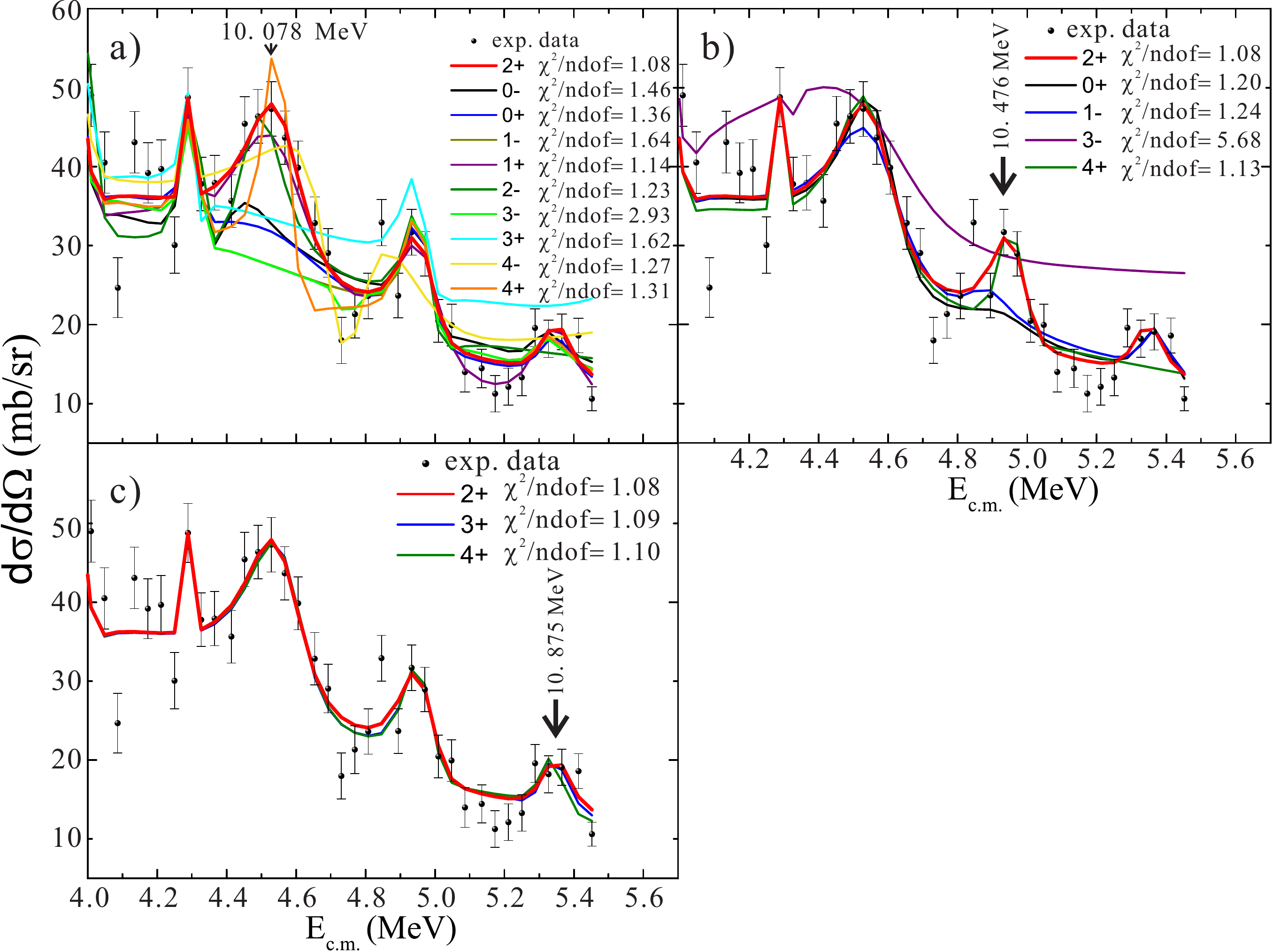}
\vspace{-3mm}
\caption{\label{fig:Jpi_test}The $J^{\pi}$ analysis of R-matrix fits for the 10.078-MeV, 10.476-MeV, and 10.875-MeV states.}
\end{center}
\end{figure}

We present here a summary of R-matrix fits for all possible spin-parity assignments to study the 10.078-MeV, 10.476-MeV, and 10.875-MeV resonant states clearly identified in this work. These three resonant states strongly contribute to the \emph{Present} $^{22}$Mg($\alpha$,$p$)$^{25}$Al reaction rate. Figure~\ref{fig:Jpi_test} illustrates the R-matrix fits for these resonant states.
 

\emph{The 10.078-MeV state} was also observed by
Shimizu \emph{et al.}~[41]
with the $^{28}$Si($^4$He,$^6$He)$^{26}$Si measurement. They determined only the resonance energy and the spin-parity remained unknown. We performed an R-matrix fit to analyze the possible $J^{\pi}$ assignments from $0^{\pm}$ to $4^{\pm}$ for the 10.078-MeV state, and found that the $J^{\pi}$=$2{^+}$ assignment yields the best-fit curve with the lowest reduced chi-square $\chi^2$/ndof, as shown in Fig.~\ref{fig:Jpi_test}\textcolor{blue}{a}. $1{^+}$ was the only assignment that cannot be fully excluded as the deviation of the fitting curve is at the edge of 1$\sigma$ confidence level, however, that assignment only modifies the $^{22}$Mg($\alpha$,$p$)$^{25}$Al reaction rate up to a factor of 0.31 for the temperature above 0.7~GK.


\emph{The 10.476-MeV state} was observed and populated also via the ($p$,$t$) reaction, which preferentially excites natural-parity states~[26]. 
Hence, we only show the R-matrix fits for natural-parity states as displayed in Fig.~\ref{fig:Jpi_test}\textcolor{blue}{b}. Fitting with $J^{\pi}$=$4^+$ can reproduce the peak around $E_x$=10.476 MeV, yet this makes the fit strongly deviated near the $2^+$ state at 10.875 MeV. Thus, we uniquely assign $J^{\pi}$=$2^+$ to the 10.476-MeV state.


\emph{The 10.875-MeV state} can only be either $2^+$, $3^+$, or $4^+$ due to the selection rule of Gamow-Teller transition according to
Thomas \emph{et al.}~[42].
The assignment of $2^+$ from the best fit may not be unique, as shown by the R-matrix fit in Fig.~\ref{fig:Jpi_test}\textcolor{blue}{c}, and thus we considered the possibility of $3^+$ or $4^+$ for this state  keeping the $^{22}$Mg($\alpha$,$p$)$^{25}$Al reaction rate for $3^+$ or $4^+$ under consideration here. This rate is shown as “alternative $J^{\pi}$=$3^+/4^+$” in Fig.~2 of the main manuscript.


\begin{figure}[b!]
\vspace{-2mm}
\begin{center}
\includegraphics[width=7.0 cm]{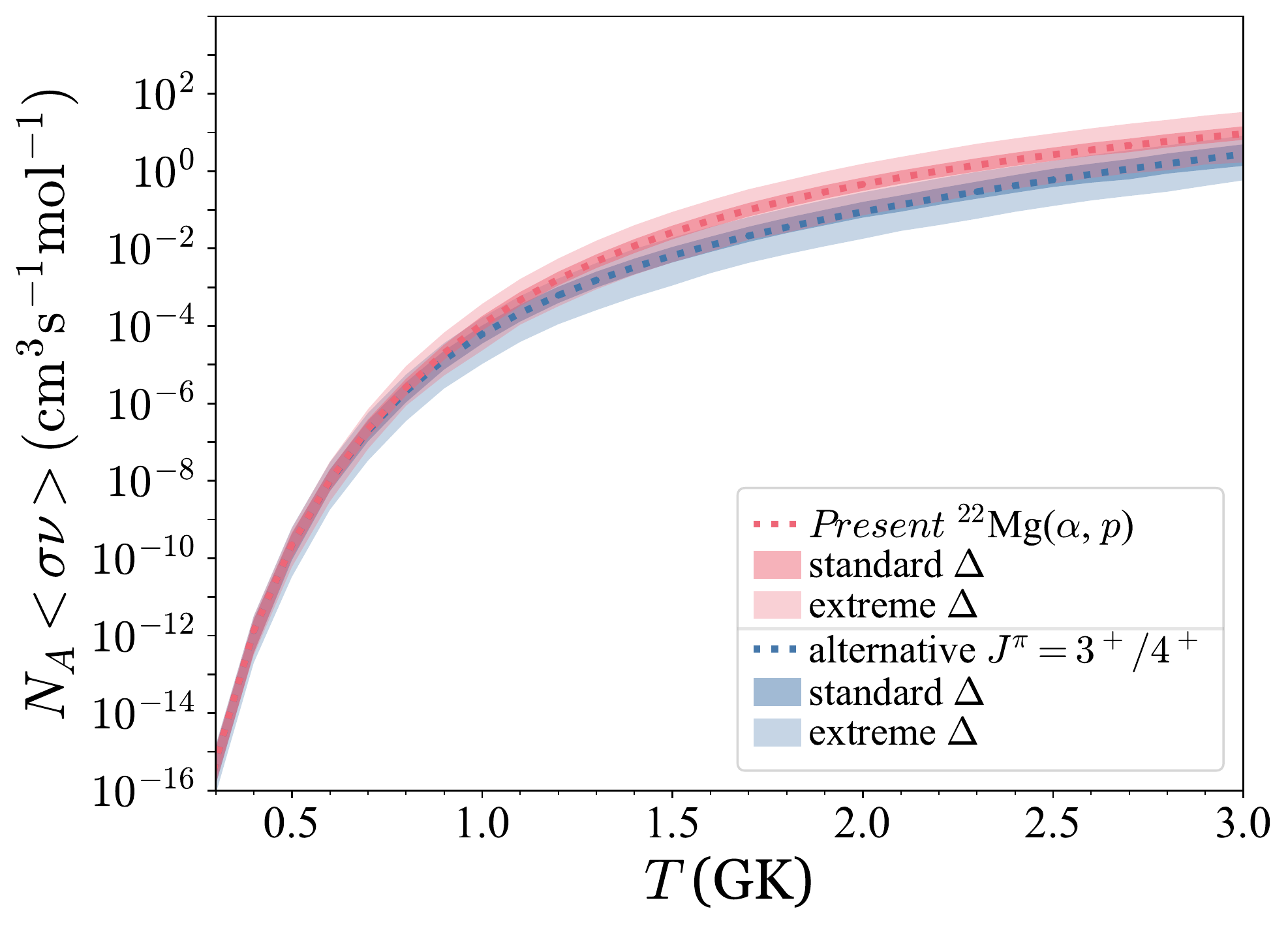}
\vspace{-3mm}
\caption{\label{fig:rate}The \emph{Present} $^{22}$Mg($\alpha$,$p$)$^{25}$Al reaction rate (red dot) and uncertainty regions based on standard (thick red zone) and extreme (light red zone) error estimations. A blue color scheme is applied for the alternative $J^\pi$=$3^+/4^+$ reaction rate.}
\end{center}
\end{figure}

\begin{center}
{\bf Uncertainty of $S_{\alpha}$ and its effect on \\ the $^{22}$Mg($\alpha$,$p$)$^{25}$Al reaction rate}
\end{center}

The $\alpha$ spectroscopic factors $S_{\alpha}$ that we have adopted here are the typical $S_{\alpha}$ values determined for the $A=26$ mirror nuclei [42, 52, 53], 
reflecting the nuclear structure information at the corresponding energy. A factor of two uncertainty for $S_{\alpha}$ is employed as the standard error based on the uncertainty estimation in a previous study for another similar system [44], 
whereas a factor of 1.30 -- 1.42 uncertainty is proposed from the root-mean-square deviation value (treated as the theoretical uncertainty) of our shell-model calculation that is based on the procedure implemented by 
Brown [54] 
using isospin non-conserving Hamiltonians of $sd$-shell nuclei [55, 56]. 
For the \emph{Present} (alternative $J^\pi$=$3^+/4^+$) rate, the thicker red (blue) zone in Fig.~\ref{fig:rate} shows the standard uncertainty for all $S_{\alpha}$. 
It is known from the previous data that $S_\alpha$ for two neighboring states with the same spin-parity may accidentally have a larger deviation, which could be even a factor of ten in some specific cases. Therefore, we also present another conservative uncertainty estimation as an extreme case, introducing a larger uncertainty of a factor of ten for all $S_{\alpha}$ values, depicted as the light red (blue) zone in Fig.~\ref{fig:rate} for the \emph{Present} (alternative $J^\pi$=$3^+/4^+$) rate. At temperature $T=1.5$ -- 1.7~GK, where the $^{22}$Mg($\alpha$,$p$) and $^{22}$Mg($p$,$\gamma$) reactions compete, the upper limit of the extreme case is a factor of 2.39 higher than the \emph{Present} rate, and the lower limit of the extreme case is only a factor of 0.83 lower than the \emph{Present} rate. Note that it is unlikely that $S_{\alpha}$ for all the states have a deviation as large as a factor of ten, and therefore we consider the uncertainty is most likely quite overestimated in this extreme case. 
Experiments to determine the $S_{\alpha}$ values in transfer reactions, e.g., $^{22}$Mg($^6$Li, $d$), and $^{22}$Mg($^7$Li, $t$), could precisely constrain the $S_{\alpha}$; however these measurements would be very difficult at the currently available worldwide facilities due to limited radioactive beam intensities. Further efforts obtaining more precise $S_{\alpha}$ would be welcomed; however, we remark that a new and more precise $^{22}$Mg($\alpha$,$p$) reaction rate is expected to be still within the range of this extreme uncertainty, and unlikely to change our conclusion.




\begin{center}
{\bf Importance of $^{22}$M\lowercase{g}($\alpha$,$p$)$^{25}$A\lowercase{l} reaction on the GS~1826$-$24 clocked burster light curve}
\end{center}

According to the sensitivity study of 
Cyburt \emph{et al.} [21] 
based on a type-I x-ray burst (XRB) model relevant to the GS~1826$-$24 clocked burster, there are other important reactions that may influence the GS~1826$-$24 XRB light curves. Here, we present further discussion with the newly deduced reaction rates of 
$^{59}$Cu($p$,$\gamma$) [57], 
$^{61}$Ga($p$,$\gamma$) [57], 
$^{14}$O($\alpha$,$p$) [58], 
$^{23}$Al($p$,$\gamma$) [59], 
and $^{18}$Ne($\alpha$,$p$) [60], 
to highlight the relevance of the $^{22}$Mg($\alpha$,$p$) reaction among them. These reactions are listed in Table~2 and Fig.~7 of Ref.~[21]. 

\begin{figure}[t!]
\begin{center}
\includegraphics[width=6.0 cm, angle=-90]{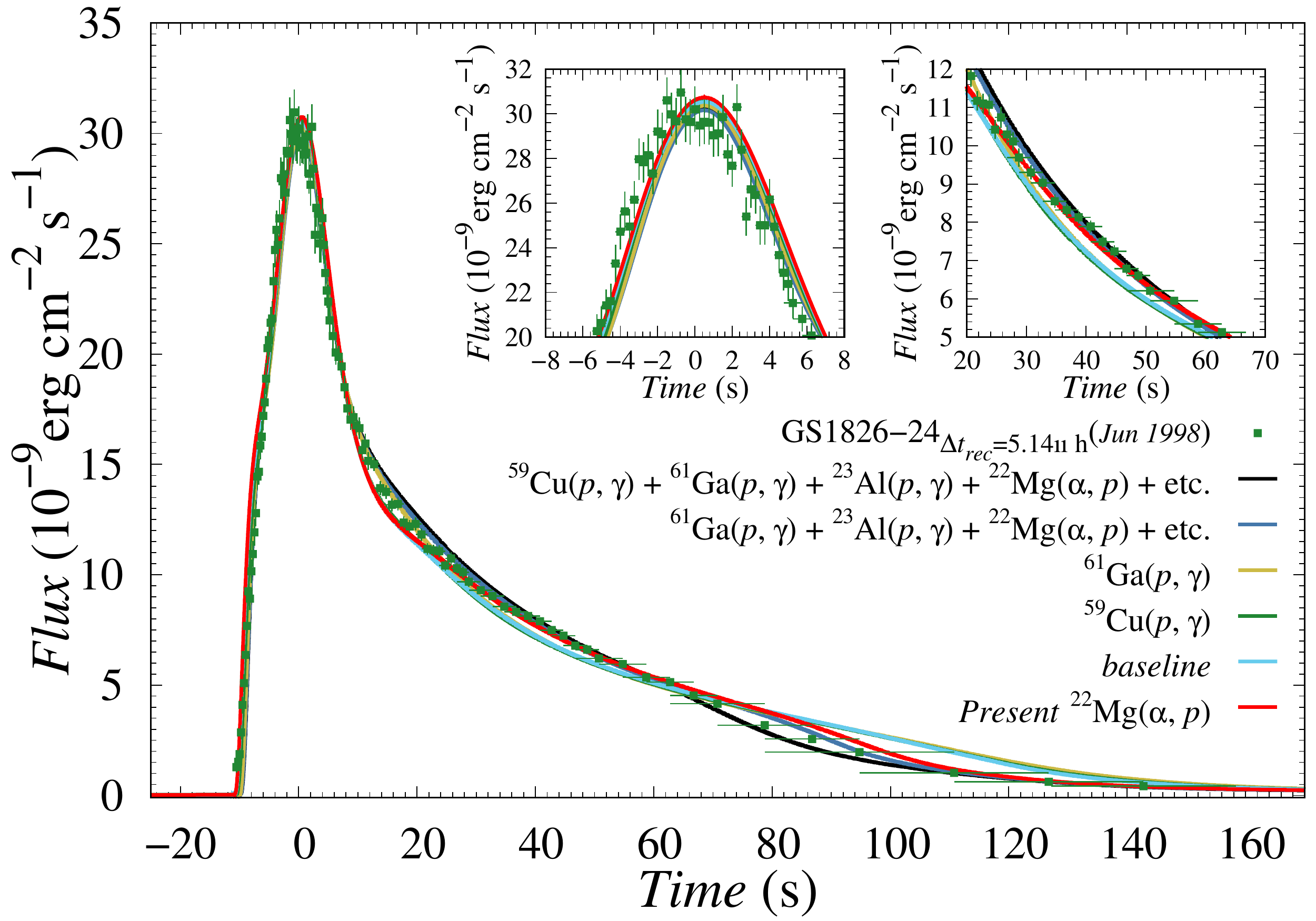}
\vspace{-3mm}
\caption{\label{fig:GS1826_Further}The best fit modeled burst light curve to the observed light curve of GS~1826$-$24 clocked burster [10, 11, 12] 
(epoch \emph{Jun 1998} [4]). 
The modeled light curves: \emph{baseline} (cyan line), \emph{Present} $^{22}$Mg($\alpha$,$p$) rate (red line), modeled burst light curves based on the newly deduced $^{61}$Ga($p$,$\gamma$) rate (yellow line), $^{59}$Cu($p$,$\gamma$) rate (green line), and the modeled light curves based on a combination of the new $^{61}$Ga($p$,$\gamma$), $^{22}$Mg($\alpha$,$p$), $^{23}$Al($p$,$\gamma$), $^{14}$O($\alpha$,$p$), $^{18}$Ne($\alpha$,$p$), $^{55}$Ni($p$,$\gamma$), $^{56}$Ni($p$,$\gamma$), $^{57}$Cu($p$,$\gamma$), $^{64}$Ge($p$,$\gamma$), and $^{65}$As($p$,$\gamma$) reaction rates (blue line), and combination of the new $^{59}$Cu($p$,$\gamma$), $^{61}$Ga($p$,$\gamma$), $^{22}$Mg($\alpha$,$p$), $^{23}$Al($p$,$\gamma$), $^{14}$O($\alpha$,$p$), $^{18}$Ne($\alpha$,$p$), $^{55}$Ni($p$,$\gamma$), $^{56}$Ni($p$,$\gamma$), $^{57}$Cu($p$,$\gamma$), $^{64}$Ge($p$,$\gamma$), and $^{65}$As($p$,$\gamma$) reaction rates (black line). The detail results will be published elsewhere by Lam \emph{et al.} [57].
}
\end{center}
\vspace{-5mm}
\end{figure}

We follow the procedure of Woosley \emph{et al.}~[47] 
to replace the statistical-model (NON-SMOKER) $^{59}$Cu($p$,$\gamma$) and $^{61}$Ga($p$,$\gamma$) reaction rates by the new shell-model $^{59}$Cu($p$,$\gamma$) and $^{61}$Ga($p$,$\gamma$) rates calculated by Lam \emph{et al.} [57] 
based on the full $pf$-model space [61]. 
The nuclear structure information provided from the shell-model calculations covers the Gamow window corresponding to the XRB temperature range. We then take these $^{59}$Cu($p$,$\gamma$) and $^{61}$Ga($p$,$\gamma$) rates prior to their publication [57], 
also the \emph{Present} $^{22}$Mg($\alpha$,$p$),
$^{14}$O($\alpha$,$p$) [58], 
$^{23}$Al($p$,$\gamma$) [59], 
$^{18}$Ne($\alpha$,$p$) [60], 
$^{55}$Ni($p$,$\gamma$) [62], 
$^{56}$Ni($p$,$\gamma$) [63], 
$^{57}$Cu($p$,$\gamma$) [64], 
$^{64}$Ge($p$,$\gamma$) [61], and 
$^{65}$As($p$,$\gamma$) [61] 
reaction rates to study the combined influence of these reactions on the GS~1826$-$24 burst light curve profile [4] 
(see Fig.~\ref{fig:GS1826_Further}).


We have studied that 
(a) the \emph{Present} $^{22}$Mg($\alpha$,$p$) reaction rate improves the modeled GS~1826$-$24 light curve to match with the observed light curve at $t=16$ -- 76~s and at $t=80$ -- 150~s (burst tail end), but not at $t=8$ -- 30~s (red line in Fig.~\ref{fig:GS1826_Further} or Fig.~3 of the main manuscript); 
(b) the new $^{61}$Ga($p$,$\gamma$) reaction rate solely improves the burst peak and at $t=8$ -- 30~s, but not at $t=30$ -- 55~s and the burst tail end (yellow line in Fig.~\ref{fig:GS1826_Further}); 
(c) the new $^{59}$Cu($p$,$\gamma$) reaction rate produces a light curve close to the \emph{baseline} light curve at $t=5$ -- 75~s, but it is over enhanced from $t=75$~s onward (green line in Fig.~\ref{fig:GS1826_Further}); 
(d) with combining the new $^{61}$Ga($p$,$\gamma$), $^{22}$Mg($\alpha$,$p$), $^{23}$Al($p$,$\gamma$), $^{14}$O($\alpha$,$p$), and $^{18}$Ne($\alpha$,$p$) reaction rates, which are the critically important reactions highlighted by Cyburt~\emph{et al.}~[21] 
and the new $^{55}$Ni($p$,$\gamma$), $^{56}$Ni($p$,$\gamma$), $^{57}$Cu($p$,$\gamma$), $^{64}$Ge($p$,$\gamma$), and $^{65}$As($p$,$\gamma$) reaction rates around the historic $^{56}$Ni and $^{64}$Ge waiting points, we remark that the modeled light curve is strikingly improved from the region of around burst peak until burst tail end (blue line in Fig.~\ref{fig:GS1826_Further}); 
(e) with combining the new $^{59}$Cu($p$,$\gamma$) with the new reaction rates in (d), we obtain that the modeled light curve is slightly enhanced at $t=10$ -- 55~s, and the light curve from $t=65$~s onward until the burst tail end is decreased and slightly lower than observation (black line in Fig.~\ref{fig:GS1826_Further}).


We have found that the role of $^{22}$Mg($\alpha$,$p$)$^{25}$Al in enhancing the light curve at $t=16$ -- 76~s and in reducing the light curve at $t=80$ -- 150~s is more decisive than other important rates mentioned above. 

\end{document}